\documentstyle[12pt]{article}
\begin{document}

\begin{titlepage}
\title{Effects of meson-exchange currents on the $(\protect\vec{e},e'p)$
structure functions}
\author{V. Van der Sluys, J.Ryckebusch\thanks{Postdoctoral research
fellow NFWO} \   and M.Waroquier\thanks{Research Director NFWO}\\
{\em  Laboratory for Nuclear
Physics}\\
{\em  Laboratory for Theoretical
Physics}\\
{\em Proeftuinstraat 86}\\
{\em 9000 Gent, Belgium}}

\vspace{.5 cm}

\date{}
\maketitle
\begin{abstract}
The response functions for the unpolarized $(e,e'p)$ and polarized
$(\vec{e},e'p)$ reaction are calculated for medium-heavy nuclei
under quasifree conditions. The formalism presented here incorporates
two-body currents related to meson-exchange and the $\Delta (1232)$
excitation. The final-state interaction of the outgoing nucleon with the
residual nucleus is handled in a HF-RPA formalism.
The sensitivity of the results to the two-body currents is discussed
for the five structure functions in quasielastic  $(\vec{e},e'p)$ scattering
off the target nuclei $^{16}O$ and $^{40}Ca$.  A selective sensitivity
to the two-body currents is obtained in the longitudinal-transverse
interference term $W_{LT}$ where two-body currents can explain part
of the discrepancy between the impulse-approximation
calculations and the data.
\end{abstract}

\end{titlepage}
\newpage
\addtolength{\baselineskip}{0.5 \baselineskip}
The coincidence $(e,e'p)$ reaction when performed under quasifree
conditions has proven to be an excellent tool in the study of
single-particle properties of the nucleus \cite{frul}. The analysis of the
quasielastic $(e,e'p)$ reaction yields information on the single-particle wave
functions, spectroscopic factors and strength distributions.
Generally, the quasielastic $(e,e'p)$ results have been analyzed within
the framework of a relativistic \cite{jin}\cite{pick} or non-relativistic
\cite{rep}\cite{jan2}\cite{bern}  distorted wave impulse
approximation (DWIA) approach. In these models the quasielastic
electromagnetic response is expected to be dominated by one-body
interactions. Hence, the nuclear current is handled in the impulse
approximation (IA) in which it will be  regarded as the sum of the
 one-body currents from the individual nucleons. The final-state interaction of
   the
outgoing proton with the residual nucleus is treated in a distorted wave
approximation either within an optical potential model \cite{rep} or within a
microscopic HF-RPA approach \cite{jan}. On the average, the $(e,e'p)$ cross
sect
   ions can be
reasonably well reproduced within this DWIA approach.

At present, improved experimental techniques have become available and
measureme
   nts
have been carried out to extract more detailed information on the response to
th
   e longitudinal
and transverse polarization states of the virtual photon.
Assuming the one-photon exchange
approximation, the unpolarized $(e,e'p)$ cross section can be written in
terms of four nuclear structure functions each multiplied with a
different kinematical factor. In case of polarized incoming electrons,
an additional
helicity $h(\pm 1)$ dependent structure function $W_{LT}'$ can be extracted
\cit
   e{ras}\cite{pick2}. The angular distribution
of the polarized $(\vec{e},e'p)$ cross section is determined by the
following expression
\begin{eqnarray}
\label{cross}
\frac{{\rm d}^{5}\sigma}{{\rm d}E_{f}{\rm d}\Omega_{E_{f}}{\rm
d}\Omega_{p_{a}}}(\vec{e},e'p)  &=&
C \left\{v_{L}W_{L} + v_{T}W_{T} + v_{TT}W_{TT}\cos 2\varphi_{a}
  \right.\nonumber \\ && \qquad\left.+ v_{LT}W_{LT}\cos\varphi_{a}  + h v_{LT}'
   W_{LT}' \sin \varphi_{a}
\right\}.
\end{eqnarray}
All five structure functions $W$ depend on the momentum and
energy transfer $(\vec{q},\omega$) of the virtual photon,
the proton momentum $p_{a}$ and proton angle $\theta_a$.
The angle between the scattering plane and the
reaction plane is denoted by $\varphi_{a}$. The various structure functions $W$
    are independent of the electron
kinematics and are sensitive in a particular way to a different aspect of the
re
   action
mechanism. A complete experimental
determination of the structure functions $W$ under suitable electron
kinematics,
 could yield additional
information on the reaction mechanism and might impose a constraint on
the theoretical model. Determination of all five structure functions
asks for polarized electrons and out-of plane experiments which are generally
difficult to realize.

 Recently, new
results of in-plane measurements have become available for
$^{16}O(e,e'p)$ \cite{saclay}\cite{chiara}\cite{wies} and
$^{40}Ca(e,e'p)$ \cite{wies}\cite{kramer}. These experiments yield information
o
   n three nuclear
structure functions $W_{L}+q^{2}/(2q_{\mu}q^{\mu}) W_{TT}$, $W_{T}$ and
$W_{LT}$
   .
In ref.\cite{saclay} the $^{16}O(e,e'p)$ data have been compared with the
results of the non-relativistic and relativistic DWIA calculations of
Van Orden {\it et al.}.  Both theoretical models are found to produce
comparable results pointing towards relativity playing a rather
unimportant role in quasielastic kinematics.
The DWIA result reproduces the shape of
the total cross section and the structure functions. However, a consistent
descr
   iption of both
the cross section and
structure functions  could neither be obtained  within the
relativistic  nor the non-relativistic approach since
the fitted reduction factors for the $^{16}O(e,e'p)^{15}N(1/2^{-},g.s)$
and the $^{16}O(e,e'p)^{15}N(3/2^{-},6.32 $ MeV) cross sections
considerably deviated from corresponding $W_{LT}$ terms.  Other
$(e,e'p)$ separation measurements on $^{16}O$ and $^{40}Ca$ have been
performed at NIKHEF~-K \cite{chiara}\cite{wies}.  These data have been
confronte
   d with the
non-relativistic DWIA model of the Pavia group \cite{rep}.  For the $(e,e'p)$
processes feeding the residual (A-1) nucleus in its ground state ($1p_{1/2}$
knockout in $^{16}O$ and $1d_{3/2}$ knockout in $^{40}Ca$) a fair agreement
between theory and experiment is observed.  For knockout from the
corresponding spin-orbit partners ($1p_{3/2}$ in $^{16}O$ and $1d_{5/2}$ in
$^{40}Ca$) the absolute $W_{LT}$ is considerably larger than what the
theory predicts.
This means that the quasielastic electron excitation of the nucleus is more
comp
   lex than it was generally believed.
{\it The aim of the present
work is to go beyond the DWIA by including two-body contributions in the
nuclear current from meson-exchange and intermediate delta excitations
and to investigate the effect of these two-body currents on the
structure functions}.

In order to determine the structure functions
the following transition matrix elements of the nuclear current
\begin{eqnarray}
\label{matr}
<J_{R}M_{R};\vec{p}_{a},1/2m_{{s}_{a}}\mid J_{\mu}(q)\mid J_{i}M_{i}>
\end{eqnarray}
have to be calculated.
The final state in this matrix element refers to the residual nucleus in a
state
$\mid \! J_{R}M_{R}\! >$ and an escaping particle with momentum $\vec{p}_{a}$
an
   d
spin projection $m_{{s}_{a}}$. Throughout this work the residual nucleus is
considered to remain in a pure hole state relative to the ground state
$\mid \!J_{i}M_{i}\!>$ of the target nucleus. The spectroscopic factor,
extracted from a least square fit of the calculated cross section to the data,
reflects the amount of hole strength in the final state.   The wave function
for
    the escaping particle
and the residual nucleus is obtained in the continuum RPA formalism as
described in ref. \cite{jan}. The RPA formalism involves a multipole expansion
in terms of linear combinations of
particle-hole and hole-particle excitations out of a correlated ground state.
As such we account for the
multi-step processes of the type depicted in Fig.~1(a).
Bound and continuum single-particle states are
taken to be eigenstates of the  HF mean-field potential
obtained with an effective interaction of the Skyrme type (SKE2) \cite{war}.
I
   n this way, we
preserve the orthogonality between the bound and the continuum states.

\begin{figure}[thb]
\vspace{7cm}
\caption[1]{{\em  Diagrams for the (e,e'p) reaction. Diagrams of the type
(a) imply virtual photon absorption on a single nucleon.
Diagrams of the type (b) (seagull term), (c) (pion-in-flight term) and (c)
(intermediate $\bigtriangleup$ creation)
refer to absorption on a two-body operator.
}}
\end{figure}

The nuclear current in the matrix element of eq. (\ref{matr})  is taken to be
 the sum of a one-body operator and a two-body operator.
The nucleonic one-body term consists of the well-known convection and
magnetization current. The two-body current is taken from a
non-relativistic reduction (retaining only terms up to the order
$1/M^{2}$ in the nucleon mass) of the lowest order Feynman diagrams with
one exchanged pion and intermediate delta excitation.
We assume pseudovector coupling of the pion to the nucleon.
This procedure gives rise to the seagull terms
(Fig. 1(b)), the pion-in-flight term (Fig. 1(c)) and
terms with
a $\Delta (1232)$ excitation in the
intermediate state (Fig. 1(d)).
In this non-relativistic approach the nuclear charge operator is not
affected by two-body contributions. The explicit expressions for the
two-body current in momentum space can be found in ref. \cite{riska}.

To account for the composite structure of the $\gamma N$, $\gamma \pi$
and $\gamma \Delta$ vertices, electromagnetic form factors have to be
introduced. For the $\gamma N$ form factor we use the common dipole
form \cite{gal}. Current conservation with the one-pion exchange potential is
me
   rely
satisfied for the seagull and the pion-in-flight current in case that the
pion ($f_{\gamma\pi}$) and
nucleon  ($f_{\gamma N}$) form factor coincide. In all further calculations we
h
   ave adopted
the $f_{\gamma\pi}$ extracted
from the vector dominance model \cite{eric}.
The delta current is divergenceless and can be multiplied with an
arbitrary form factor without violating the charge-current conservation rules.
For simplicity, we assumed that $f_{\gamma\Delta}=f_{\gamma N}$ in all
calculations presented here.
The short-range structure of the $\pi NN$ and $\pi N
\Delta$  vertices
is implemented in a phenomenological way by introducing hadronic form factors.
A
   s is usually done
the monopole form is adopted. For both types of vertices the same pion
cut-off mass $\Lambda_{\pi}(=\Lambda_{\pi NN}=\Lambda_{\pi \Delta N}$)
is used.

The final state wave function in the
matrix element of eq. (\ref{matr}) is evaluated using a multipole expansion in
t
   erms of
distorted waves.
So, the nuclear current operator is  decomposed in the well-known electric and
m
   agnetic
transition operators $T^{el}_{JM}$ and $T^{mag}_{JM}$ \cite{for}.
If we restrict ourselves
 to the evaluation of
 diagrams of the type depicted in Fig.~1,  reduced
matrix elements of the following type
\begin{eqnarray}
\label{trans}
<0^{+}\mid\mid T_{J}^{(1)}(q)+T_{J}^{(2)}(q)\mid\mid (ph^{-1});J>
\end{eqnarray}
remain to be calculated.
In this expression $T_{J}^{(1)}$ and $T_{J}^{(2)}$ refer to the IA and
the pionic contribution.
The two-body part of the transition operators is handled exactly and involves
tw
   o active nucleons in the
absorption process. Hence, in the evaluation of (\ref{trans}), the
two-body part has been expressed in terms of two-body matrix elements.
The explicit expression for these two-body matrix elements for the
diagrams of Fig.~1 can be found in ref. \cite{jan3}.

All results presented here are obtained  under the kinematical
conditions of the $^{16}O(e,e'p)$ \cite{chiara}\cite{wies}
and  $^{40}Ca(e,e'p)$ \cite{wies}\cite{kramer} NIKHEF~-K experiments. These
expe
   riments were
performed under perpendicular kinematics at  ($\omega=96$ MeV, $q=460$
MeV) ($^{16}O$)
 and ($\omega=116$ MeV, $q=446$ MeV) ($^{40}Ca$).
The structure functions are plotted as a function of the missing
momentum $p_{m}=\mid \vec{p}_{a}-\vec{q} \mid$.

\begin{figure}[thb]
\vspace{8.5cm}
\caption[1]{{\em Structure functions for $^{16}O(e,e'p)$ from the $1p_{3/2}$
orbit including one (dotted line) and two-body contributions in the nuclear
current.
(a) $W_{LT}$ for
 different values of the pion cut-off mass in the hadronic
form factor. Dash-dotted line:
$\Lambda_{\pi} = 650$ MeV; dashed line: $\Lambda_{\pi} = 1200$ MeV; full line:
$
   \Lambda_{\pi} =
800$ MeV.
(b) $W_{T}$ considering different expressions for the pion form factor. Dashed
line: $f_{\gamma\pi}=f_{\gamma N}$; full line:
$f_{\gamma\pi}$ from vector dominance model. No spectroscopic factors are
consid
   ered.}}
\end{figure}

Some theoretical uncertainties exist with respect to the pion cut-off
mass $\Lambda_{\pi}$ in the
hadronic form factor and to the electromagnetic pion form factor.
Before confronting our theoretical approach with the data, we investigate
the sensitivity of the results to these parameters.
 Whereas the Bonn potential leads to a cut-off mass
$\Lambda_{\pi} = 1200$ MeV \cite{mach}, recent studies on the triton system
\cite{sasa} seem to prefer a
smaller value ($\Lambda_{\pi} = 810$ MeV).  We have performed calculations
including all diagrams depicted in Fig.~1 with different values
of $\Lambda_{\pi}$. As a representative example  we display in
Fig.~2(a) the
longitudinal-transverse interference structure function $W_{LT}$ for proton
knockout out of the $1p_{3/2}$ orbit
in $^{16}O(e,e'p)$.
Results are displayed for a pion cut-off mass that should
be considered as a lower limit ($\Lambda_{\pi} = 650$ MeV) and an upper limit
($
   \Lambda_{\pi} = 1200$
MeV). Also shown is the $W_{LT}$ structure function as
 obtained within the IA . The uncertainty of the results due to the
theoretical ambiguity in  $\Lambda_{\pi}$
should be estimated around 20\% of the total two-body contribution. In all
furth
   er calculations we used a
cut-off mass $\Lambda_{\pi}=800$ MeV . With this lower
cut-off value heavier mesons such as the $\rho$-meson are partially
taken into account \cite{riska}.

The sensitivity of the present approach to the $f_{\gamma\pi}$ form factor
in the pion-in-flight current is investigated in
Fig.~2(b). In the forthcoming discussion it will be shown that, of all
structure functions, the $W_{T}$ exhibits the largest sensitivity to the
pion-in
   -flight term.
We have plotted $W_{T}$ for
proton knockout out of the $1p_{3/2}$ orbit in $^{16}O$. The
results including the full nuclear current are compared with the IA
predictions.
Two different $f_{\gamma\pi}$ form factors are considered in the calculations.
Firstly, the pion form factor is set equal to the nucleon form factor
and, as a consequence, current conservation is satisfied. Secondly, we adopt
the $f_{\gamma\pi}$ form factor as derived from the vector dominance model.
Clearly, the results are rather insensitive to the choice of the pion
form factor.

\begin{figure}[thb]
\vspace{12 cm}
\caption[1]{{\em Structure functions $W_{T}$ and $W_{LT}$ for proton knockout
of
   f $^{16}O$
from the $1p_{1/2}$ and $1p_{3/2}$
orbit including one (dotted line) and two-body contributions in the nuclear
curr
   ent.
Dash-dotted line: one-body current and seagull current; dashed
line:one-body, seagull and pion-in-flight current; full line: one-body,
seagull, pion-in-flight and delta current . The data are taken from ref.
\cite{chiara}. The curves are multiplied with a spectroscopic factor
($S(1p_{1/2})=0.60$, $S(1p_{3/2})=0.51$).
}}
\end{figure}
\begin{figure}[thb]
\vspace{12cm}
\caption[1]{{\em $W'_{LT}$ for the $(\vec{e},e'p)$ reaction
off $^{16}O$ from the
$1p_{1/2}$ and $1p_{3/2}$
orbit including (a) only one-body contributions in the nuclear current (dotted
l
   ine)
 and (b) both one and two-body contributions in the nuclear current
(full line). No spectroscopic factors are included.
}}
\end{figure}

The impact of the different components in the nuclear current on the  structure
   functions
in the $^{16}O(\vec{e},e'p)$ reaction is illustrated in
Figs.~3 and ~4.
In Fig.~3 we display the transverse   and interference structure functions
$W_{T
   }$ and
$W_{LT}$ for
$^{16}O(e,e'p)^{15}N(1p_{1/2}^{-1},g.s.)$ and
$^{16}O(e,e'p)^{15}N(1p_{3/2}^{-1},6.32 MeV)$. All curves are multiplied with a
   spectroscopic factor
extracted from a best fit of the calculated cross section to the data.
In conformity with the results
of refs. \cite{rep}\cite{wies} we do not arrive at a {\it simultaneous}
descript
   ion
of the total cross section and the $W_{LT}$ structure function in
the IA. Furthermore, a spin-orbit dependence of the
results is observed: whereas the IA seems to work reasonably well for the
$1p_{1/2}$ state, the absolute value of  $W_{LT}$ is severely
underestimated for the $1p_{3/2}$ orbit.
{}From Fig.~3 it is clear
that for both single-particle states the seagull and pion-in-flight current
exhibit the same characteristics. Whereas the seagull contribution enhances
the transverse and charge-current interference response functions, the
pion-in-flight current has the opposite effect. Generally, the
pion-in-flight term has a smaller influence on the structure functions than
the seagull current.
 The spin-dependent behaviour of the $W_{LT}$ term for the two states
originates
from the contribution of the $\Delta(1232)$ current. For the $1p_{1/2}$ state a
   further quenching is observed
in contrast with the $1p_{3/2}$ results where the deviation from the
IA approach becomes more pronounced. As can be noticed, the discrepancy between
experiment and the DWIA results for $W_{LT}$ of the $1p_{3/2}$ state can
be partially ascribed to the two-body contributions in the nuclear current.

In contrast to the cross section for which the effect
of the pionic currents is hardly visible, two-body contributions in the
nuclear current cannot be discarded in order to obtain a complete description
of
    the
structure functions.

In Fig.~4 we display the fifth structure function $W'_{LT}$ as can be extracted
from the polarized $^{16}O(\vec{e},e'p)$ reaction for the same orbits. The
calcu
   lations are performed
using the same kinematics as before. To our knowledge, no data of this
response function are available as yet. In the near future a large
number of polarization experiments will be carried out at MIT-Bates \cite{pap}.
   From a theoretical standpoint, this additional
measurable quantity $W_{LT}'$ has as its main advantage that it vanishes
identic
   ally in
the plane wave impulse approximation (the escaping
proton is described by a plane wave) \cite{ras} . So, the
shape of the calculated $W_{LT}'$ term is {\it completely determined} by the
fin
   al-state
interaction and the different contributions in the nuclear current.
In comparison with the results for the $W_{LT}$ term, we observe a similar
spin-dependent behaviour with respect to the two-body nuclear current.
For both single-particle orbits, the shape remains almost unaffected compared
to
    the
DWIA calculation.
Whereas a small quenching
is observed for the $1p_{1/2}$ state, pionic
contributions seem to enhance the fifth structure function for the
$1p_{3/2}$ state.
Generally, the effect of the two-body contributions is more apparent in
the $W_{LT}$ term than in the corresponding $W_{LT}'$ structure function.

\begin{figure}[thb]
\vspace{12cm}
\caption[1]{{\em $W_{LT}$ for $^{40}Ca(e,e'p)^{39}K(1d_{3/2}^{-1};g.s.)$ and
$^{40}Ca(e,e'p)^{39}K(1d_{5/2}^{-1}; 5.26$ MeV, $5.60$ MeV, $ 6.36$ MeV).
(a) dashed line: one-body current,
(b) full line: one and two-body contributions in the nuclear current.
An overall spectroscopic factor is used for the cross sections and the
corresponding structure functions (S($1d_{3/2}$)=0.48, S($1d_{5/2}$)=0.33).
The data are taken from ref. \cite{wies}.}}
\end{figure}

Until now, all conclusions drawn referred to $^{16}O$ results.
Similar calculations have been performed for proton ejection from the
target nucleus $^{40}Ca$. We have calculated cross sections and
structure functions for the spin-orbit partners $1d_{3/2}$ and
$1d_{5/2}$. The results for the
interference structure function $W_{LT}$ are
displayed in Fig.~5.
Once again, meson contributions in the current are not negligible and can
accoun
   t for
part of the discrepancy between the IA calculations and the data.
Due to the spin-dependent behaviour of the $\Delta$ current,
the $W_{LT}$ function for the $1d_{3/2}$ state and the $1d_{5/2}$ state is
modif
   ied in
a different way.
This conclusion is in conformity with the results for the spin-orbit partners
$1
   p_{1/2}$ and $1p_{3/2}$ in
$^{16}O$.

Summarizing, the results presented here indicate the importance of
two-body contributions in the nuclear current in order to reach a complete
descr
   iption
of the $(\vec{e},e'p)$ cross section and structure functions under quasifree
con
   ditions.
Going beyond the impulse approximation, we have accounted for two-body
contributions  in the current related to one-pion exchange and intermediate
$\De
   lta(1232)$
creation. Calculations were performed for proton knockout
from the target nuclei $^{16}O$ and $^{40}Ca$. The results were shown to
be rather insensitive to the model assumptions with respect to the pion
and hadronic form factor. The charge-current interference
structure function $W_{LT}$ is found to be strongly affected by the two-body
cur
   rents.
A rather good agreement with the data could be obtained within a model
that accounts for the FSI within an HF-RPA model and in which one and
two-body photoabsorption mechanisms are included.
Rather than the cross section the separated structure functions are
sensitive to the different aspects of the reaction mechanism. In this
sense, a further exploration of the separate structure functions opens good
pers
   pectives
to obtain a better
insight into the nature of the $(e,e'p)$ reaction mechanism.

{\bf Acknowledgement}

The authors are grateful to K.Heyde for fruitful discussions and suggestions.
This work has been supported by the Inter-University Institute for
Nuclear Sciences (IIKW) and the National Fund for Scientific Research (NFWO).

\end{document}